\documentclass[12pt]{iopart}
\usepackage{xcolor}
\usepackage{graphicx}
\usepackage{dcolumn}
\usepackage{bm}
\usepackage{multirow}
\usepackage[numbers,sort&compress,square]{natbib}
\usepackage{hyperref}
\usepackage{subcaption}
\usepackage{verbatim}
\usepackage{soul}
\usepackage{iopams} 
\usepackage{acronym}
\usepackage[mathlines,running]{lineno}
\usepackage{tabularx}
\usepackage{listings}

\newcommand{\LSU}{Department of Physics, Louisiana State University, 202 Nicholson Hall
Baton Rouge, LA 70803, USA}
\newcommand{\MIT}{LIGO Lab, Massachusetts Institute of Technology, Cambridge, MA 02139, USA}
\newcommand{\LLO}{LIGO Livingston Observatory, Livingston, LA 70754, USA}
\newcommand{\UNO}{Department of Physics, University of Omaha Nebraska, 6001 Dodge Street Durham Science Center, Omaha, NE 68182, USA }

\begin{document}
\title{Noise in the LIGO Livingston Gravitational Wave Observatory due to Trains}
\author{J~Glanzer$^{1}$, 
S~Soni$^{2}$, 
J~Spoon$^{3}$, 
A~Effler$^{4}$, 
G~Gonz\'alez$^{1}$}

\address{$^1$\LSU}
\address{$^2$\MIT}
\address{$^3$\UNO}
\address{$^4$\LLO}

\vspace{10pt}
\begin{indented}
\item[]
\end{indented}

\begin{abstract}
Environmental seismic disturbances limit the sensitivity of LIGO gravitational wave detectors. 
Trains near the LIGO Livingston detector produce low frequency ($0.5$--$10~\mathrm{Hz}$) ground noise that couples into the gravitational wave sensitive frequency band ($10$--$100~\mathrm{Hz}$) through light reflected in mirrors and other surfaces. 
We investigate the effect of trains during the Advanced LIGO third observing run, and propose a method to search for narrow band seismic frequencies responsible for contributing to increases in scattered light. Through the use of the linear regression tool Lasso (least absolute shrinkage and selection operator) and glitch correlations, 
we identify the most common seismic frequencies that correlate with increases in detector noise as $0.6$--$0.8~\mathrm{Hz}$, $1.7$--$1.9~\mathrm{Hz}$, $1.8$--$2.0~\mathrm{Hz}$, and $2.3$--$2.5~\mathrm{Hz}$ in the LIGO Livingston corner station.   
\end{abstract}

\submitto{\CQG}

\section{Introduction}
The detection of gravitational waves from extreme astrophysical events has opened the door to exciting discoveries. The Advanced LIGO~\cite{TheLIGOScientific:2014jea} and Advanced VIRGO~\cite{TheVirgo:2014hva} detectors have detected many gravitational waves from coalescing binaries of black holes and neutron stars~\cite{LIGOScientific:2021djp}. 
Since the first detection in 2015, there has been an influx of more events detected. By the end of the first observing run (O1)~\cite{LIGOScientific:2016dsl}, both the LIGO Scientific Collaboration (LSC) and the LIGO Virgo and KAGRA (LVK) Collaboration had reported three binary black hole gravitational wave events. The second observing run (O2)~\cite{LIGOScientific:2018mvr} not only detected seven binary black hole mergers, but the first merger of two neutron stars. In the third observing run (O3)~\cite{LIGOScientific:2021djp}, a total of 79 events were detected with improved sensitivity due to increased laser input power and the introduction of squeezed light techniques ~\cite{Buikema:2020dlj, LIGOScientific:2021usb}. 
These observations require 
measurements of differences in arm lengths of the order of $\rm 10^{-21}\,m/\sqrt{Hz}$
at $150~\mathrm{Hz}$.

Short duration non-astrophysical transients known as ``glitches" adversely impact the detector's data quality and complicate the process of identification of gravitational waves in the data ~\cite{LIGOScientific:2017tza, LIGOScientific:2019hgc, dcc:worstoffenders_sidd}. Environmental and instrumental noise are common sources of these brief transients. Earthquakes, ocean waves, trains, and human activity can ``shake" the detector. Quadruple suspension systems and passive/active seismic isolation are used to dampen the effects. The in-vacuum optical tables incorporate an active vibration isolation system providing attenuation of environmental seismic noise below $1~\mathrm{Hz}$. Above $1~\mathrm{Hz}$, the quadruple pendulum optic suspensions provide passive isolation in the horizontal and vertical degrees of freedom. \cite{2015CQGra..32r5003M}. 
Although the ground motion is largest at low frequencies (below 10 Hz), it produces transients in gravitational wave strain data in the $10$--$150~\mathrm{Hz}$ range 
due to non-linear coupling. At the LIGO Livingston Observatory (LLO), 50$\%$ of the noise transients in the LLO detector in O3 were due to light reflected in mirrors and other surfaces (``scattered" or ``stray" light) \cite{Glanzer_2023}.

In this paper, we investigate scattering noise 
caused by trains passing near 
LLO. Noise from trains can enter the interferometer by causing displacements of particular scattering surfaces in the detector. In Section \ref{seismic activity}, we introduce how scattered light produces noise in the detector. In Section \ref{Methods}, we describe the methods used to investigate the effect of trains in producing transient noise in the detector. 
In Section \ref{results}, we present the results of our analysis showing the advantages of the methods used, and discuss possible surfaces producing the scattering glitches.

\section{Seismic activity and scattered light}\label{seismic activity}
The motion of the ground where a gravitational wave detector is located can lead to bad data quality. Earthquakes and wind shake the instrument in $0.03$--$0.1~\mathrm{Hz}$ band, ocean currents in the Gulf of Mexico are the main source of increased microseismic motion in $0.1$--$0.3~\mathrm{Hz}$ band and human activities such as logging,  construction work and vehicles can cause increased ground motion in $1$--$6~\mathrm{Hz}$ anthropogenic band. Although all of these frequency bands are not in the sensitive detector gravitational wave band (10Hz-4 kHz), ground motion can couple non-linearly into that band. For example, during O3 transients due to scattered light were the most frequent source of transient noise at both LLO and LHO (LIGO Hanford Observatory), and were a result of increased ground motion in one or more of these seismic bands. 

\begin{figure}
    \centering
    \includegraphics[width=0.95\textwidth]{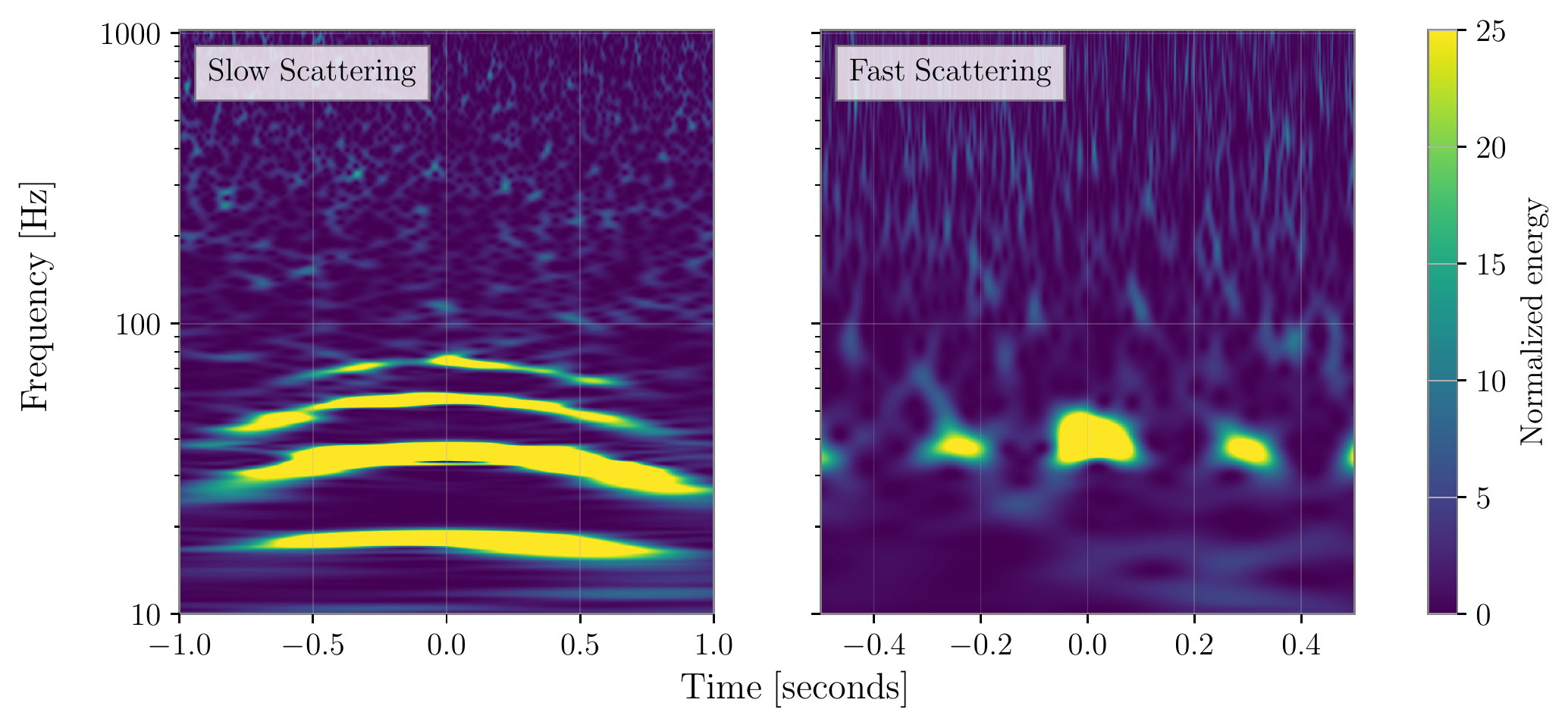}
    \caption{Time-frequency spectrograms known as Q-scans of the different types of scatter in the main gravitational wave data channel. Slow scatter has multiple arches, indicating several light reflects. Higher frequency harmonics are not present for fast scatter.  }
    \label{fig:slow_fast}
\end{figure}

Light  from the main laser path is scattered by a mirror and can be reflected by another surface. A fraction of this scattered light can rejoin the main path, and introduce  a time dependent phase modulation, shown in eq \ref{eq_phase}, to the phase of the laser field. The additional phase shows up as noise $h_{\rm{ph}}$ in the detector data, shown in eq \ref{eq_ph_noise}.  

\begin{equation}
  h_{\rm{ph}}(f) = \frac{K}{2}\frac{\lambda}{2\pi L}\mathcal{F}[\sin{\delta \phi}] \label{eq_ph_noise}
\end{equation}
where

\begin{equation}
    \phi(t) = \phi_{0} + \delta \phi_{\rm{sc}}(t) \label{eq_phase} = \frac{4 \pi}{\lambda}[x_{0} + \delta x_{\rm{sc}}(t)],
\end{equation}
here, $K$ is the ratio of stray light amplitude to the amplitude of light in the main beam (usually unknown but very small), $\lambda$ is laser wavelength (1064 nm) and $L$ is the length of interferometer arms (4km). $\mathcal{F}$ indicates a Fourier transform, $x_{0}$ is the static path which corresponds to $\phi_{0}$, and $\delta x_{\rm{sc}}$ is the time-dependent displacement of the scattering surface which gives rise to the additional phase $\delta \phi_{\rm{sc}}$. If the phase modulation is small, $\sin\phi\approx\phi$ and the noise couples linearly; if $\delta x_{\rm{sc}}$ is a large fraction of $\lambda$, the noise couples non-linearly. 
In this latter case, as a result of fringe wrapping, the phase noise $h_{\rm{ph}}(f)$ associated with scattered light can show up as arches in $h(t)$ spectrogram. If we differentiate the phase $\phi = 2\pi f t$  in eq \ref{eq_phase} with respect to time, we obtain the frequency of the noise:
\begin{equation}
    f(t) = |\frac{2v_{\rm{sc}}(t)}{\lambda} |\label{peak_f}
\end{equation}
where $v_{\rm{sc}}$ is the velocity of the scatterer and $f$ is the peak frequency of the transients. If the surface excited by ground motion is in approximately periodic motion, the noise will appear as arches in a time-frequency spectrogram shown in the left plot of Fig \ref{fig:slow_fast}.  From Eq \ref{peak_f} we can see that a scatterer moving with higher velocity will lead to transients at higher peak frequency. 
A scatterer receiving or reflecting too little light or moving with small velocity amplitude will cause transients below the gravitational wave band \cite{Soni:2020rbu}. 

Transient scattered noise can be classified into two main categories, depending on the frequency of the ground motion producing it: ``slow scattering" and ``fast scattering", shown in 
fig \ref{fig:slow_fast}. Slow scattering is usually a result of increased ground motion in earthquake and microseismic band, whereas fast scattering transients are usually a result of high ground motion in the anthropogenic band. In this paper we focus on fast scattering transients caused by increased anthropogenic motion due to trains \cite{Soni:2021cjy}. (For details on slow scattering and its reduction during O3, we refer to \cite{Soni:2020rbu}.) 

\section{Methods}\label{Methods}

\subsection{Detector seismic couplings}
Physical environment monitoring (PEM) is a system of sensors monitoring and recording aspects of the physical environment surrounding the interferometer, allowing studies of the impacts of environmental noise on the detector. One methodology used to find the potential location and intensity of noise coupling in the detector is to inject a known form of disturbance into the detector and study the differential arm length (DARM) response. These injections, known as PEM injections include magnetic, seismic and acoustic injections. The site of injection, its amplitude, the frequency band, etc are varied to study the impact on DARM of these changing parameters. Vibrational seismic injections have been regularly performed at LLO and LHO to characterize the coupling between ground motion in anthropogenic band and detector hardware \cite{Nguyen:2021ybi, Effler_2015}. The seismic disturbances are injected using vibrational shakers and monitored using accelerometers. If a detector component in the vicinity of the injection location has stray light incident on it, and has a resonance that falls within the frequency band of the injection, then its excitation would make noise in DARM.

\begin{table}[h]
\centering
\setlength{\tabcolsep}{6pt}
    \begin{tabular}{|c|c|} 
 \multicolumn{1}{c}{Detector component} & \multicolumn{1}{c}{Resonance frequencies}\\ 
 \hline
ITMY Cryobaffle  & $3.82$ Hz, $4.19$ Hz \cite{alog:itmy_res}\\
\hline
End  Y Cryobaffle & $3.49$ Hz, $4.62$ Hz \cite{alog:etmy_res_a, alog:etmy_res_b}\\
\hline
End  X Cryobaffle & $4.10$ Hz \cite{alog:etmx_res_a} \\
\hline
End Y Arm Cavity Baffle & $1.6$ Hz \cite{alog:etmy_acb_a} \\
\hline
ITMY Arm Cavity Baffle & $1.5$ Hz 
\cite{alog:etmy_acb_a_comment} \\
\hline
\end{tabular}
    \caption{Resonant frequencies of various cryobaffles and arm cavity baffles.}
    \label{tab:resonance_shaker} 
\end{table}

Baffles are an example of detector hardware designed for the purpose of absorbing and/or redirecting any incident stray light on them. These components are installed at multiple locations in the detector. Depending on the surface, location and geometry of the baffle, they can backscatter stray light towards the test mass mirror and into the main beam \cite{dcc:baffle_backscatter}. The injections carried out since the end of O3 have revealed the presence of resonances in multiple baffles across the detector. At these resonant frequencies, the baffles can inject scattered light noise back into the gravitational wave readout up to as high as 100 Hz \cite{alog:etmy_acb_a}. Table \ref{tab:resonance_shaker} lists the dominant cryobaffle and arm cavity baffle resonances found at LLO during shaker injection tests in 2020 and 2022. These resonances have been mechanically damped \cite{alog:acb_damped}.

\subsection{Identifying Fast Scattering Glitches}

During O3, Fast Scattering glitches were the most common glitch type at LLO. Fast scatter shows up as short duration arches in the time-frequency spectrograms of the primary GW channel.

\begin{figure}[ht]
\captionsetup[subfigure]{font=scriptsize,labelfont=scriptsize}
   \centering
    \begin{subfigure}[b]{1.0\textwidth}
        \centering
         \includegraphics[width= \textwidth]{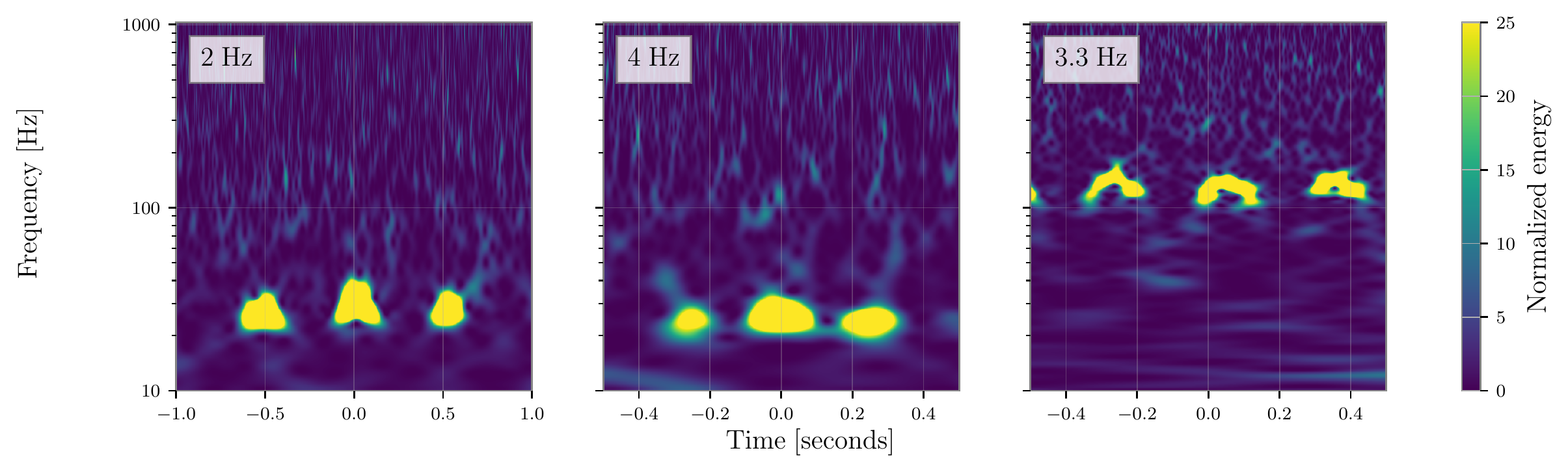}
         
         \label{fig:fastscatter}
    \end{subfigure}
    \caption{Time-frequency spectrograms of the most common types of fast scatter observed during and post O3. The different types are characterized by how often the arches repeat \cite{soni_2021}. }
\end{figure}

About 27$\%$ of all glitches classified by Gravity Spy with a confidence of 90$\%$ were Fast Scattering at LLO. Fast Scattering glitches have been observed to occur when there is an increase in ground motion, specifically, when there is an increase in the microseismic frequency range ($0.1$--$0.3~\mathrm{Hz}$) and the anthropogenic frequency range ($1$--$6~\mathrm{Hz}$). Ocean waves and currents in the Gulf of Mexico increase microseismic activity, whereas construction work and logging, thunderstorms, wind, and trains near the detector site leads to an increase in anthropogenic seismic activity at LLO \cite{Daw:2004qd, Soni:2020rbu}. 

Several different types of Fast Scattering have been observed, characterized by the frequency of their repeating arches. The most common types seen during O3 are referred to as $4$ Hz followed by $2$ Hz fast scatter. In the data taken between O3 and O4, $3.3$ Hz fast scatter was observed, with a higher peak frequency than was observed during O3.

\subsection{Identifying trains and characterizing the ground motion they produce}

There are several detector characterization tools used to detect, characterize and classify noise transients. These tools include but are not limited to Omicron, GWpy Omega Scans, and GravitySpy ~\cite{Davis:2021ecd,hveto,gwdetchar,gwpy,Robinet:2020lbf,Essick:2020qpo, Zevin:2016qwy}. Omicron is an event trigger generator (ETG) which is used to search for excess power in the gravitational wave data. These transients detected by omicron are colloquially known as Omicron triggers or just triggers. Each of these triggers is assigned parameters such as event time, frequency, duration, signal to noise ratio (SNR) etc. Omicron triggers are further used downstream by a number of other detector characterization tools including GravitySpy. GravitySpy is an image classifier based on Convolutional Neural Networks (CNN), used to classify transient noise into different categories or classes depending on the time-frequency morphology of the noise ~\cite{coughlin_scott_2021_5649212}. 

\begin{figure}[ht]
\captionsetup[subfigure]{font=scriptsize,labelfont=scriptsize}
   \centering
    \begin{subfigure}[b]{0.7\textwidth}
        \centering
         \includegraphics[width= \textwidth]{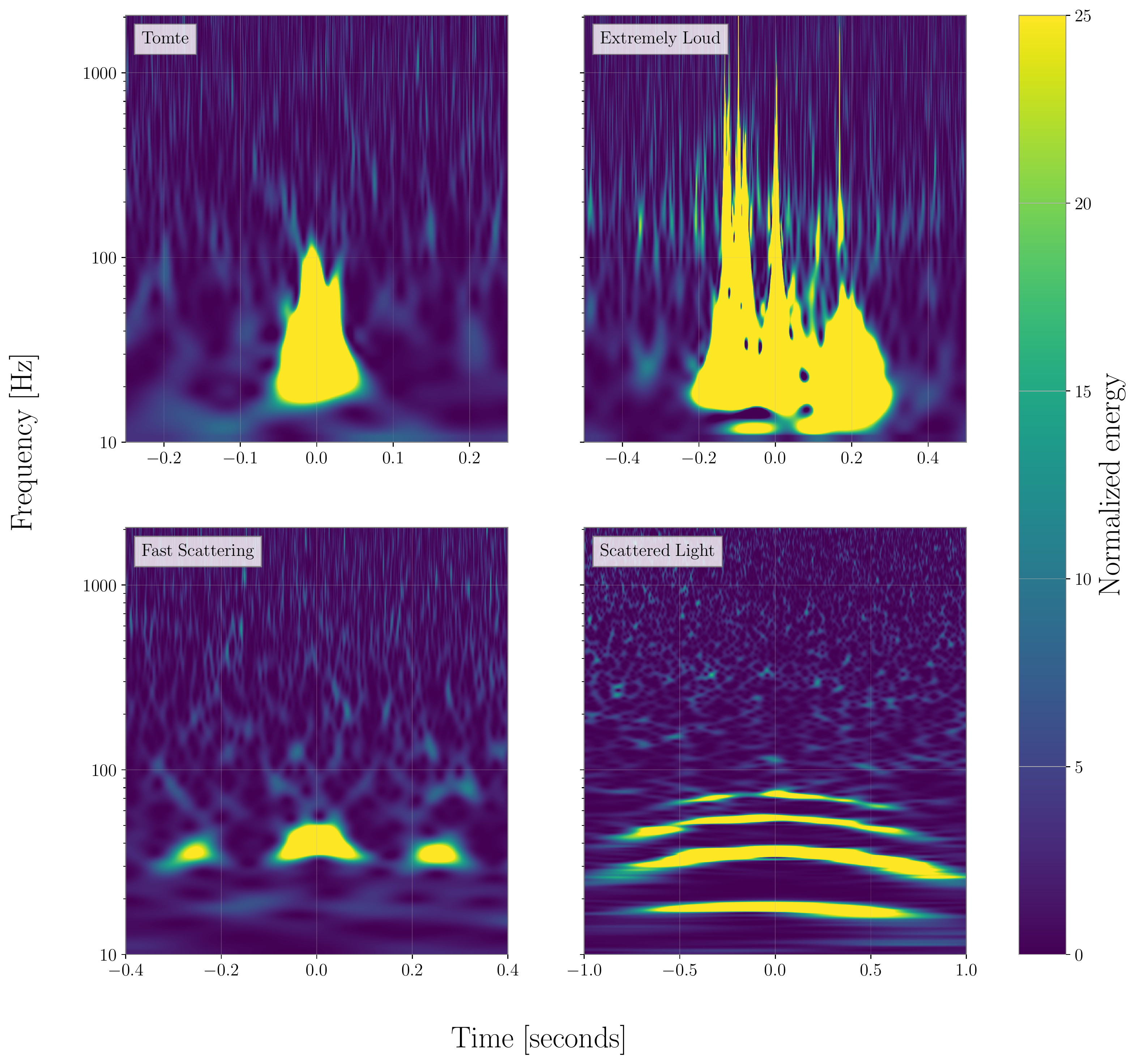}
         
         \label{fig:gspyglitches}
    \end{subfigure}
    \caption{Examples of different glitch classifications from Gravity Spy. Different glitches are visualized via omega scans in time-frequency space.}
\end{figure}

\begin{figure}[ht]
\captionsetup[subfigure]{font=scriptsize,labelfont=scriptsize}
   \centering
    \begin{subfigure}[b]{0.5\textwidth}
        \centering
         \includegraphics[width= \textwidth]{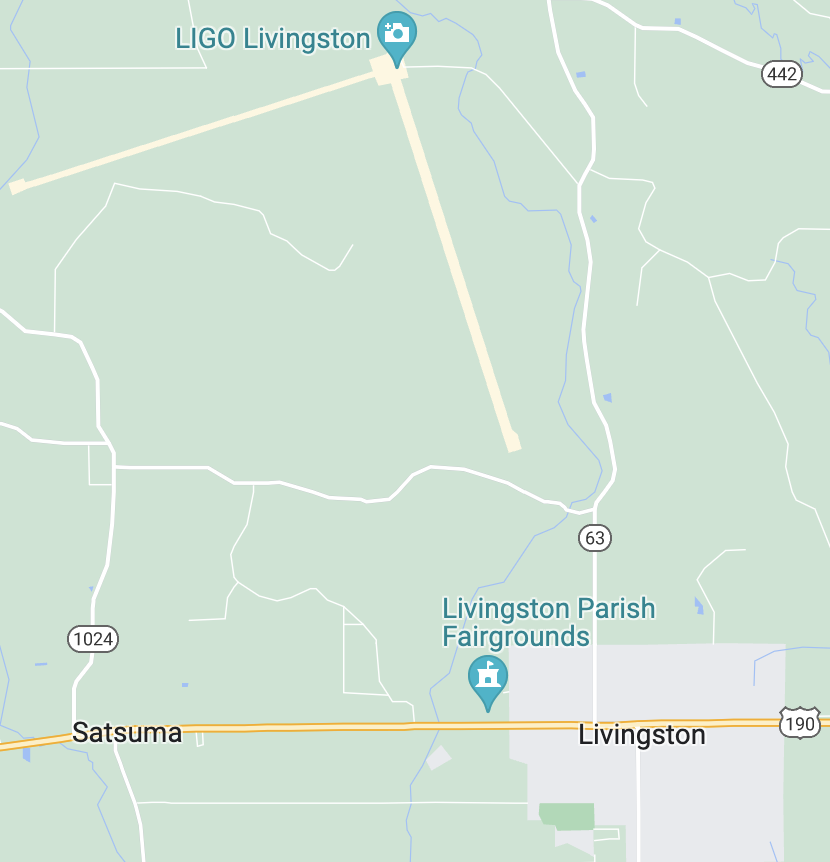}

    \end{subfigure}
    \caption{Location of LLO relative to the train track in Livingston, Louisiana. The track runs parallel to highway 190, which passes through Livingston. The end of the $Y$ arm is approximately two miles from the track. \emph{Image: Google Maps}}
    \label{fig:livingstonligopic}
\end{figure}

One of the major sources of increased noise in the anthropogenic frequency range is trains passing by the LLO Y end station, see Figure \ref{fig:livingstonligopic}. We use data from 3 seismometers (Guralp\textsuperscript{\textregistered} CMG-3T \cite{2021CQGra..38n5001N}) at LLO, one located at each end station and one in the corner station. Each seismometer measures ground motion in the $X$, $Y$, and $Z$ directions. 

We developed a PYTHON tool in an effort to define, systematically, a cut on the root mean square ground motion above which we will consider a train to be affecting the interferometer. The tool finds the peaks and widths of trains based on a specified ground motion threshold. For our analysis, ground motion larger than a 600 nm/s threshold in ETMY$\_$Y $1$--$3~\mathrm{Hz}$  for more than 2.5 minutes were considered trains. (This is typically caused by trains, but if there is any other source producing large motion in the same band, it will have a similar effect to noise produced by trains.) This identified a total of 791 trains in O3. 

\begin{figure}
    \centering
    \includegraphics[width=0.95\textwidth]{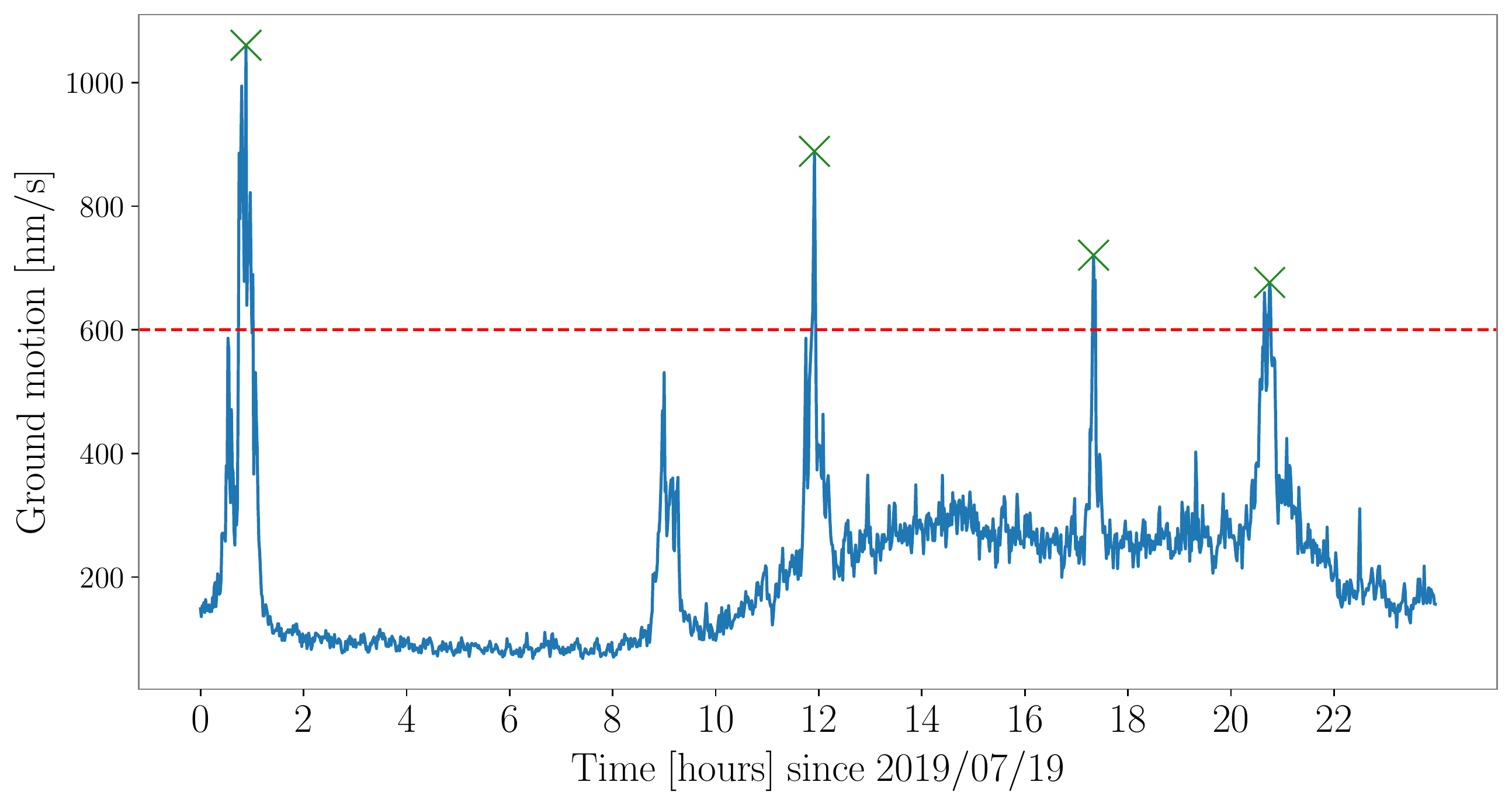}
    \caption{Example output of the python script used to select trains via seismometer data from the $Y$ channel of ETMY in the $1$--$3~\mathrm{Hz}$ frequency range. The horizontal bar is the cut above which data is associated with a train.}
    \label{fig:train_script}
\end{figure}

Earthquakes and other seismic activity may occur during the trains. In order to investigate the effect of solely due to seismic noise due to trains, we also set cuts on earthquake and microseismic activity. For earthquakes, the cut was $>100~\mathrm{nm/s}$ in ITMY$\_$X, and for high microseismic activity the cut was $>1500~\mathrm{nm/s}$ in ITMY$\_$X.
In O3, there were 199 trains that did not occur during either an earthquake, high microseismic activity, or already high anthropogenic noise. (During the day time hours near LLO, the anthropogenic noise sometimes increases due to activities such as logging and construction work.) The cut for what was considered already high anthropogenic motion was $>200~\mathrm{nm/s}$ in ETMY$\_$Y. We applied this limit approximately 30 minutes before each train to catch if there was a rise in the motion levels.

\begin{figure}[!tph]
  \begin{subfigure}[b]{0.48\textwidth}
    \includegraphics[width=\textwidth]{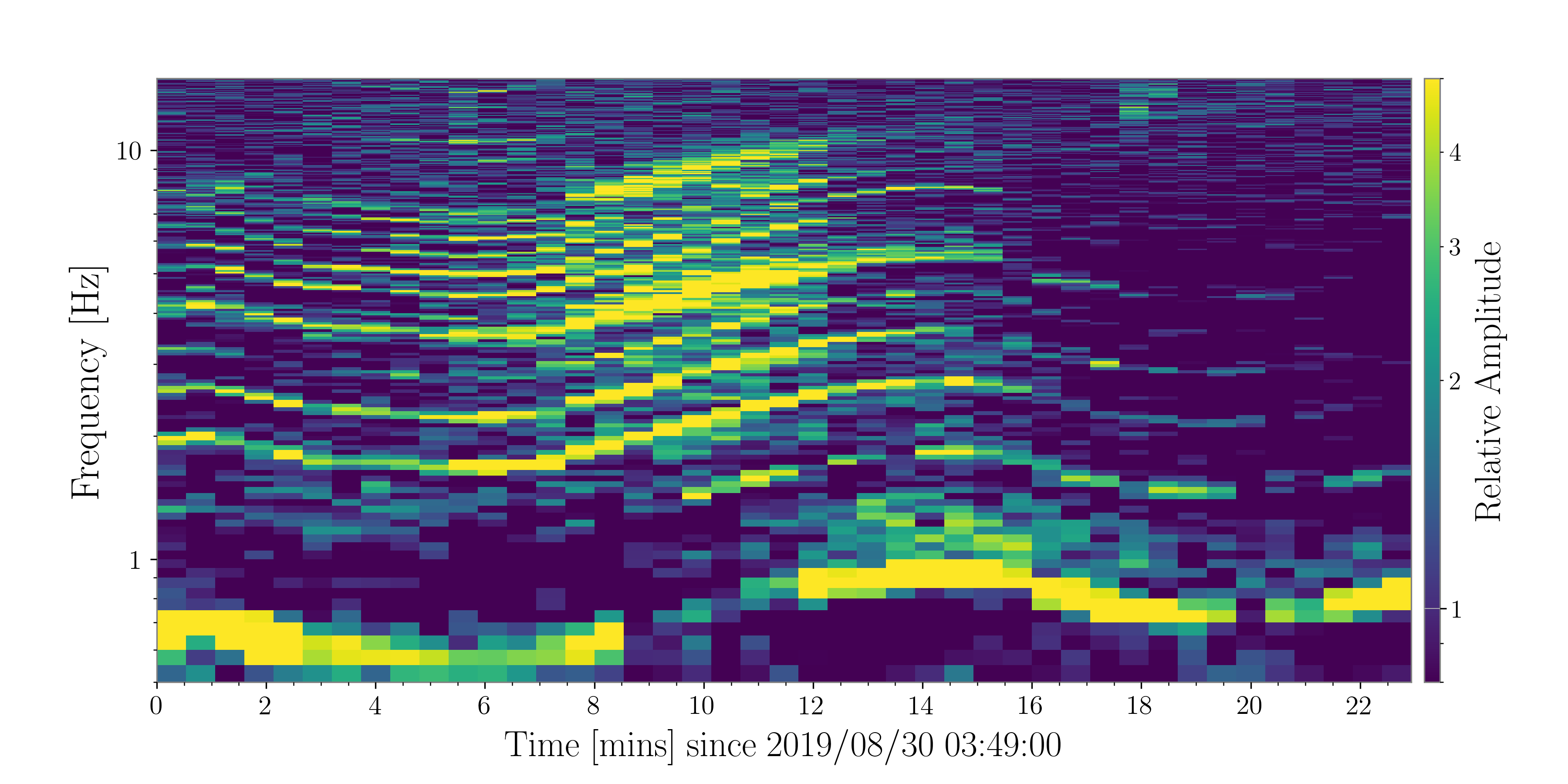}
    \caption{Spectrogram of data from $Y$ channel of the seismometer at ETMY}
    \label{fig:emtyyspectrogram}
  \end{subfigure}
  \hfill
  \begin{subfigure}[b]{0.48\textwidth}
    \includegraphics[width=\textwidth]{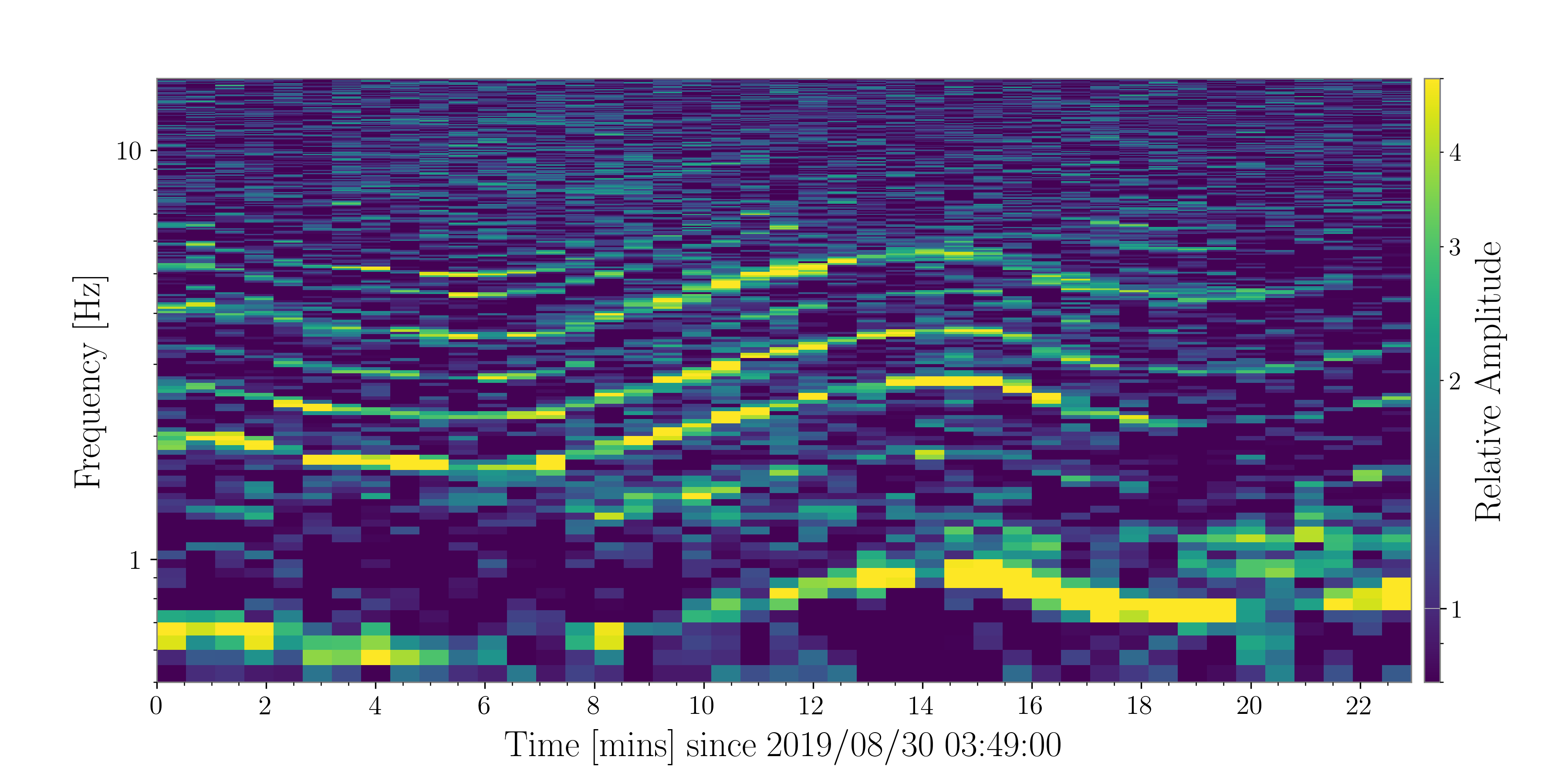}
    \caption{Spectrogram of data from $Y$ channel of the seismometer at ITMX}
    \label{fig:itmxyspectrogram}
  \end{subfigure}
  \caption{Visible harmonics during a train in O3. The fundamental frequency can be seen towards the lower end, ranging from $0.5$--$0.7~\mathrm{Hz}$. While there are more visible lines at higher frequencies near the Y end station as that is closer to the train track, the frequencies are identical in all stations. }
  \label{fig:train}
\end{figure}

To characterize the ground motion produced by trains, we produced time-frequency spectrograms of data acquired from the seismometers; see Figure \ref{fig:train}. There are harmonic lines with changing frequency in the spectrograms. The harmonics can be explained by the physical structure of train wagons ~\cite{10.1785/0220170092}. Train wagons distribute their weight along four axles, connected in pairs of two to the bogies on each wagon end. The load of each train axle results in a periodic force on the ground. The frequency of this periodic source depends on the geometric makeup of the axles and the train speed. The repeated force of the axles on the ground is the cause for the spectral line spacing in the spectrograms. 
The train accelerating/decelerating produces smoothly changing frequencies (the Doppler effect is negligible). For trains, the fundamental harmonic typically lies around $0.5$--$1.0~\mathrm{Hz}$. If we take the typical cargo train wagon to be approximately 18 meters, this suggests a train speed of $9$--$18$ meters per second. In Figure \ref{fig:emtyyspectrogram} and \ref{fig:itmxyspectrogram}, the fundamental harmonics are around $0.7~\mathrm{Hz}$.

Although the amplitude and frequency of the ground motion changes smoothly during the time of a train, which is visible in the seismometers, the noise in the gravitational wave readout shows several different short ``bursts" of increased amplitude. We suspect each burst is produced by the seismic noise exciting mechanical resonances in different scattering surfaces. 

\begin{figure}[ht]
\captionsetup[subfigure]{font=scriptsize,labelfont=scriptsize}
   \centering
    \begin{subfigure}[b]{0.9\textwidth}
        \centering
         \includegraphics[width= \textwidth]{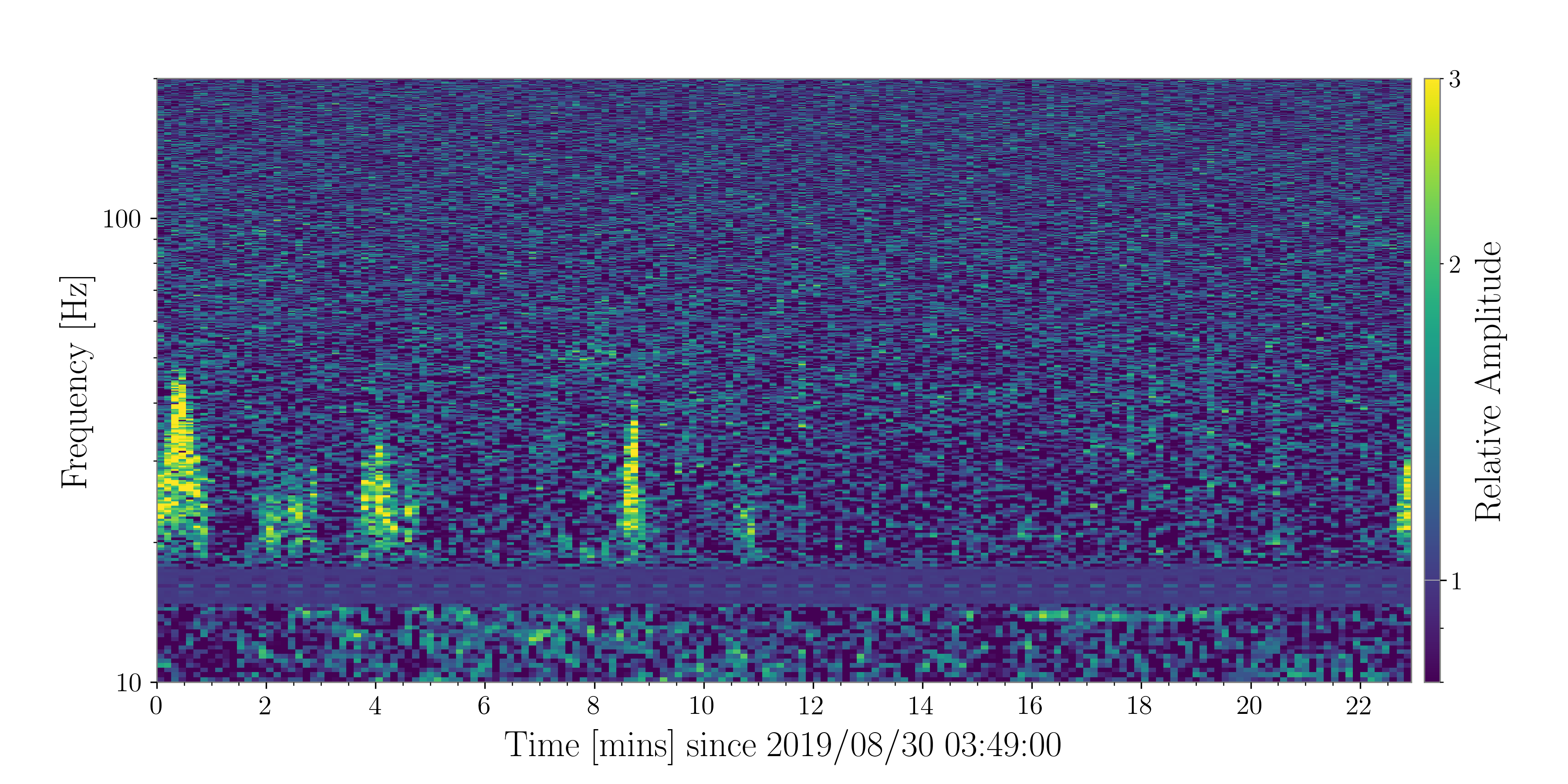}
         
         \label{fig:darmspec}
    \end{subfigure}
    \caption{Spectrogram of increased detector noise during an O3a train on 2019-08-30. Bursts of increased noise are observed in the main GW data channel.}
\end{figure}


\subsubsection{Lasso Correlation Analysis}

To find correlations between the calibrated strain data and data from the seismometers, we used the least absolute shrinkage and selection operator regression method, Lasso \cite{10.2307/2346178}. Lasso is a regularization technique that adds together a linear combination of a data set to create a best fit model of some specified data. In doing so, Lasso uses a shrinkage method in which the coefficients of determination are shrunk towards zero, and the less important features of the combination of channels from the data set are omitted by shrinking their respective coefficients to zero. Lasso uses a parameter called alpha, which is what determines how many of the coefficients are driven to zero \cite{Walker_2018}. We selected an alpha value of 0.003, and restricted the coefficients to be positive. This helps prevent over-fitting of the data, allowing us the ability to find physical meaning of correlations between increased seismic noise and strain noise. 

For our analysis, we use data from the seismometers to create a model of the strain noise during the time of trains. We bandpass the seismic data from $0.3$--$10~\mathrm{Hz}$ in steps of 0.2 Hz,
and then create a time series with the rms of each band with a time step of 5 seconds. This gives us approximately ~300 narrow seismic bands to correlate with DARM for each seismometer and direction. From this, we are able to look at specific band-passed frequencies in a particular station as well as direction where we can further analyze how they fit with the calibrated strain data. With these parameters, we have approximately 30 - 40 narrow seismic bands that correlate with the strain noise per train. 

To confirm which bands were the most frequently correlated with the gravitational wave channel, we identified a control group of times in which there was no noise in the detector for comparison. We selected 59 quiet times near trains (within $\pm$ 1 hour), with durations of 15 minutes and ran Lasso on them. During these quiet times, we wouldn't expect any bursts of power (due to the trains) and therefore we can use this to check the validity of Lasso correlations during trains.

\begin{figure}[ht]
\captionsetup[subfigure]{font=scriptsize,labelfont=scriptsize}
   \centering
    \begin{subfigure}[b]{0.7\textwidth}
        \centering
         \includegraphics[width= \textwidth]{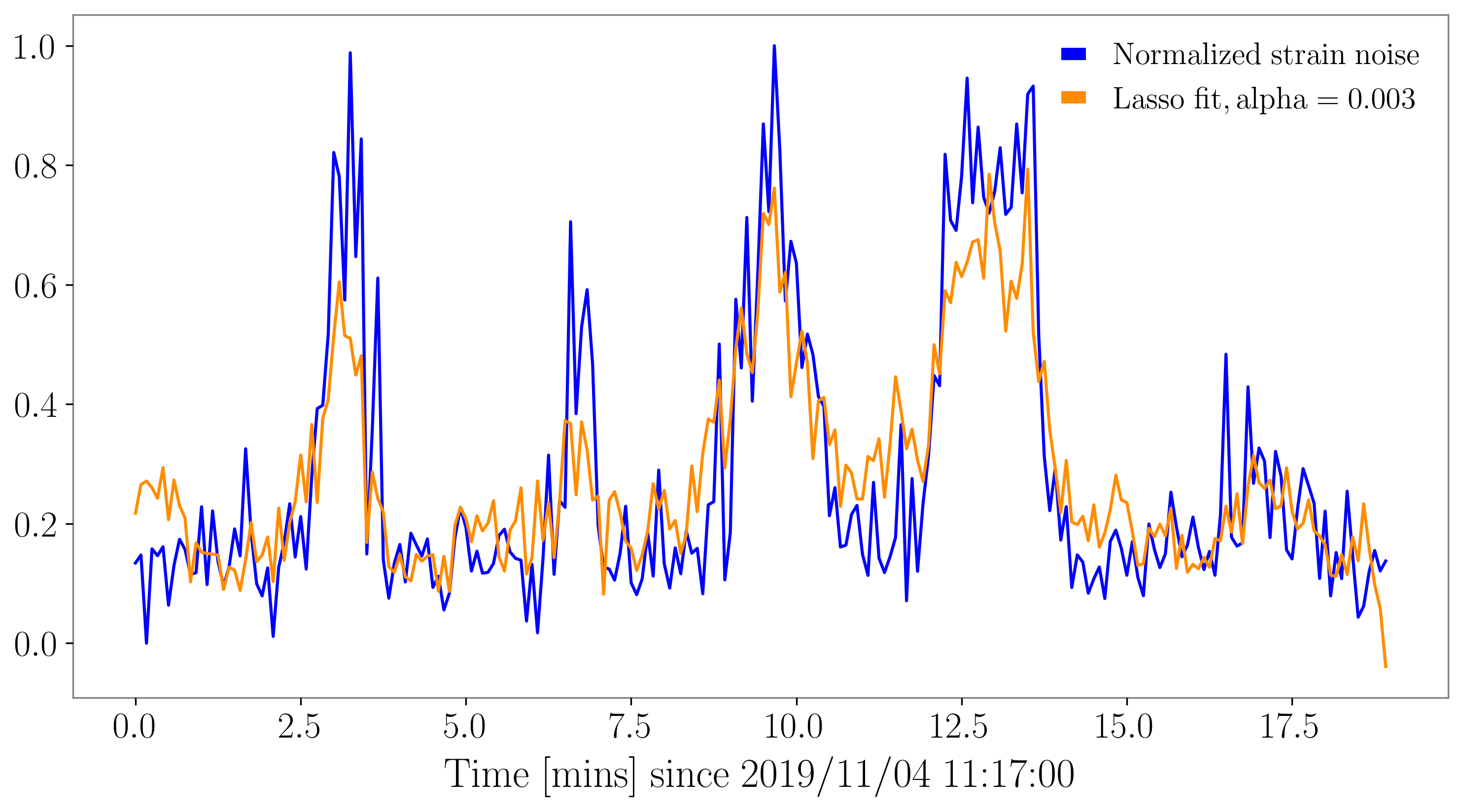}
         
         \label{fig:lassofitexample}
    \end{subfigure}
    \caption{An example Lasso fit, with an r-square value of 0.72 for train 2019-11-04 11:17:00 UTC. In blue, we have the normalized strain noise which serves as our primary model. The orange represents a linear combination of data from all three seismometers. }
\end{figure}

\subsubsection{Spearman Correlation Analysis}
To study the relationship between train induced anthropogenic ground motion and transient noise in the gravitational wave readout, we developed an algorithm that calculates the Spearman correlation between the two variables of interest \cite{boslaugh_watters_2009}. These variables are ground motion velocity measured in $\mathrm{nm/s}$ and the transient noise rate. We use Spearman correlation because we assume the relationship between how much the ground is moving and the rate of transients is monotonic, i.e. if one increases so does the other and vice versa. Days with higher ground motion for example, are associated with increased rate of transients \cite{Soni:2020rbu}.  The exact relationship between the ground motion and transient noise rate is complicated and depends on multiple factors such as how much of the ground motion translates to scatterer motion, the amount of scattered light hitting the scatterer etc. 


For each train, we start with raw ground motion velocity data sampled at $512~\mathrm{Hz}$ in the $X$, $Y$ and $Z$ directions recorded by the 3 seismometers located at End X , End Y and corner station. We then band pass this timeseries data in frequency bands from $0.3~\mathrm{Hz}$  to $4.8~\mathrm{Hz}$  Hz in steps of $0.3~\mathrm{Hz}$ and calculate the root-mean-square of 30 seconds of data. Next, we calculate the rate of transient noise for each of these trains, by calculating the total number of Omicron triggers between the start and end time of the train, and dividing that by the duration of the train. For these triggers, we apply the SNR threshold ($5 < \rho < 50$) and lower and upper frequency cut-offs at $10~\mathrm{Hz}$  and $200~\mathrm{Hz}$  respectively. For any given train, we have a total of 135 time-series streams, since the data from 3 seismometer locations along 3 axes is bandpassed in 15 frequency bands. Next, we calculate the Spearman correlation coefficient between the median ground motion value of these 135 time-series and the transient noise rate.  

From the collection of 791 trains we randomly selected 330 trains and ran the algorithm to calculate the correlation coefficients defined above. This gives us, for each train, 135 coefficients, each one showing the strength of monotonic relationship between ground motion in a particular band and the location and axis of that ground motion in the detector. We then calculate the median of these coefficients, across all the trains analyzed. The results are shown in Table \ref{tab:O3_trains_rate_motion}.

\section{Results}\label{results}
\subsection{O3 Lasso Results}
With Lasso, we can see which narrow seismic frequencies are the most correlated with increased detector noise during the time of trains. In particular, we want to determine which frequencies and detector locations are the most common in an effort to locate potential scattering surfaces. As described in section \ref{Methods}, 199 trains during O3 were analyzed with Lasso.

\begin{figure}[ht]
\captionsetup[subfigure]{font=scriptsize,labelfont=scriptsize}
   \centering
    \begin{subfigure}[b]{0.9\textwidth}
        \centering
         \includegraphics[width= \textwidth]{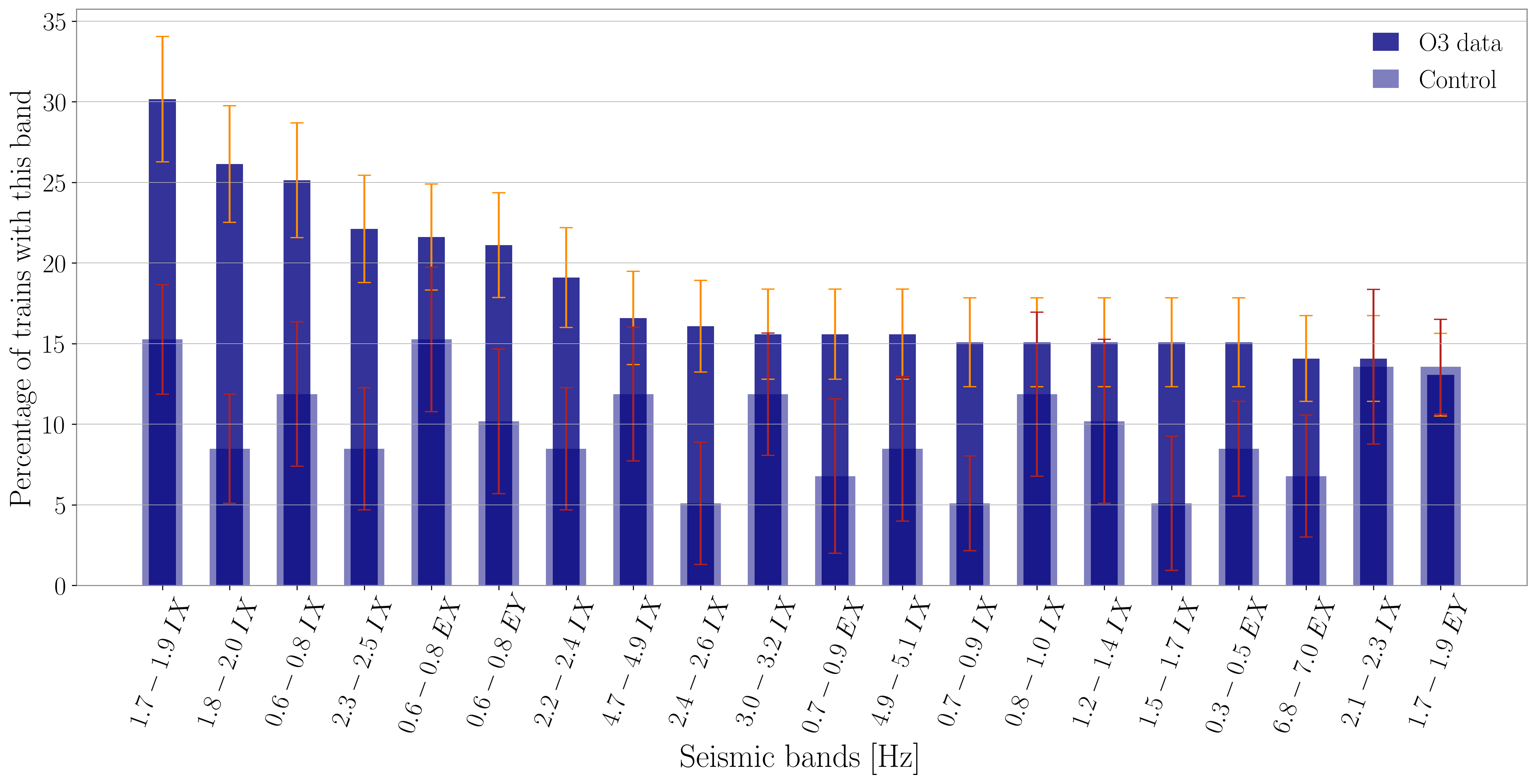}

    \end{subfigure}
    \caption{Lasso results displaying the top 20 most common seismic frequencies that have correlations with increased strain noise during trains in O3. The corner station (ITMX) is the most common detector location.}
    \label{fig:o3lassoresults-station}
\end{figure}

\begin{figure}[ht]
\captionsetup[subfigure]{font=scriptsize,labelfont=scriptsize}
   \centering
    \begin{subfigure}[b]{0.9\textwidth}
        \centering
         \includegraphics[width= \textwidth]{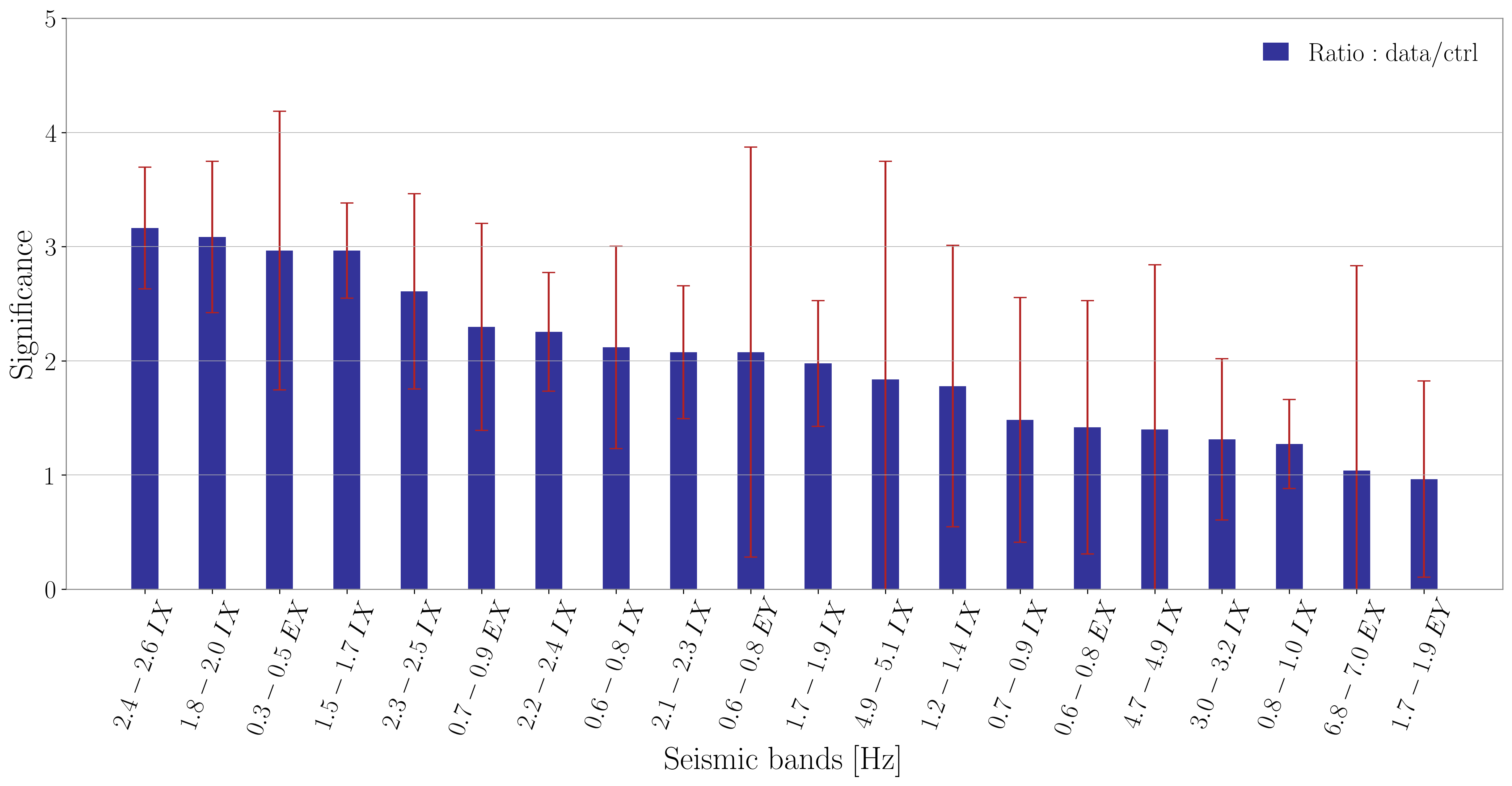}

    \end{subfigure}
    \caption{Significance of each seismic band as compared to the control group.}
    \label{fig:o3ratioplot}
\end{figure}

\begin{figure}[ht]
\captionsetup[subfigure]{font=scriptsize,labelfont=scriptsize}
   \centering
    \begin{subfigure}[b]{0.9\textwidth}
        \centering
         \includegraphics[width= \textwidth]{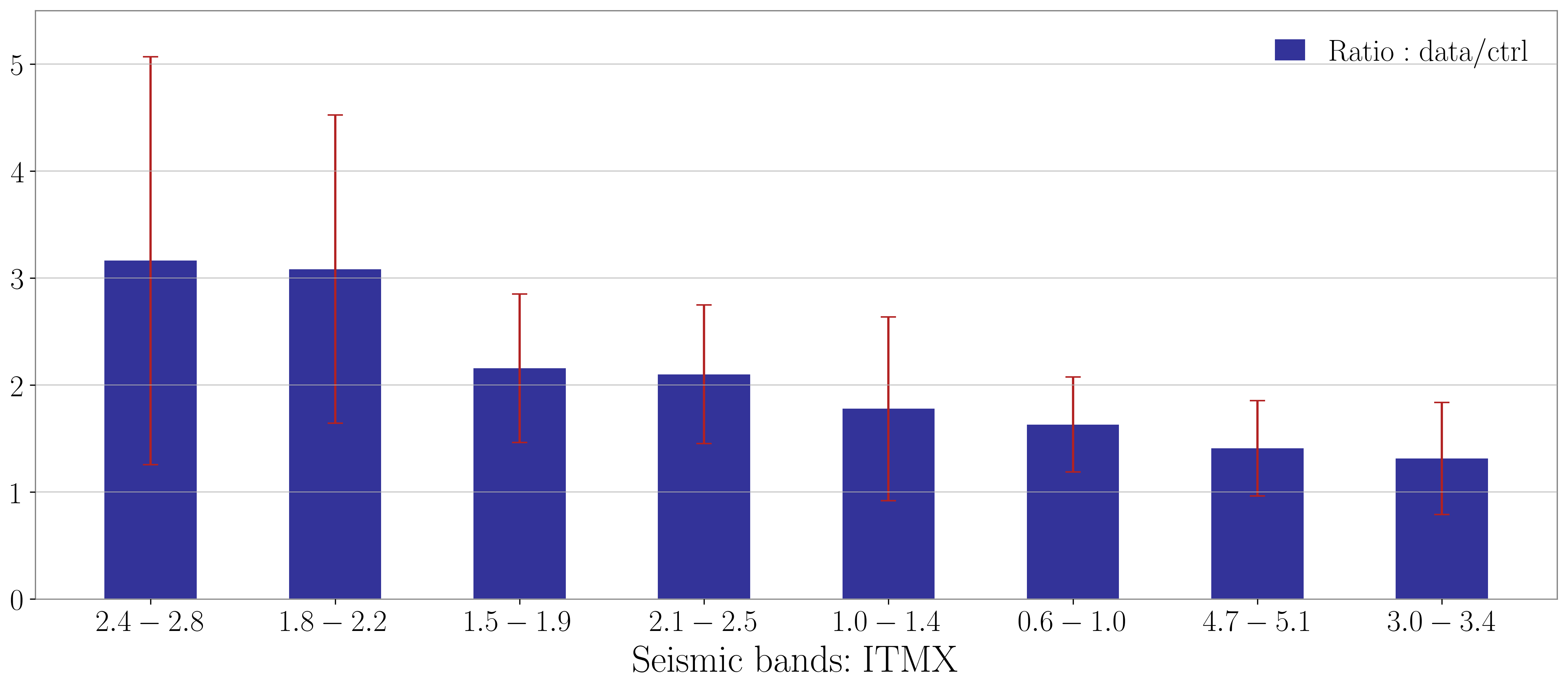}

    \end{subfigure}
    \caption{Significance of wider frequency grouping of only the ITMX Lasso data as seen in Figure \ref{fig:o3ratioplot}. }
    \label{fig:o3lassoresults-grouped}
\end{figure}

\begin{figure}[!tph]
\captionsetup[subfigure]{font=scriptsize,labelfont=scriptsize}
   \centering
    \begin{subfigure}[b]{1.0\textwidth}
        \centering
         \includegraphics[width= \textwidth]{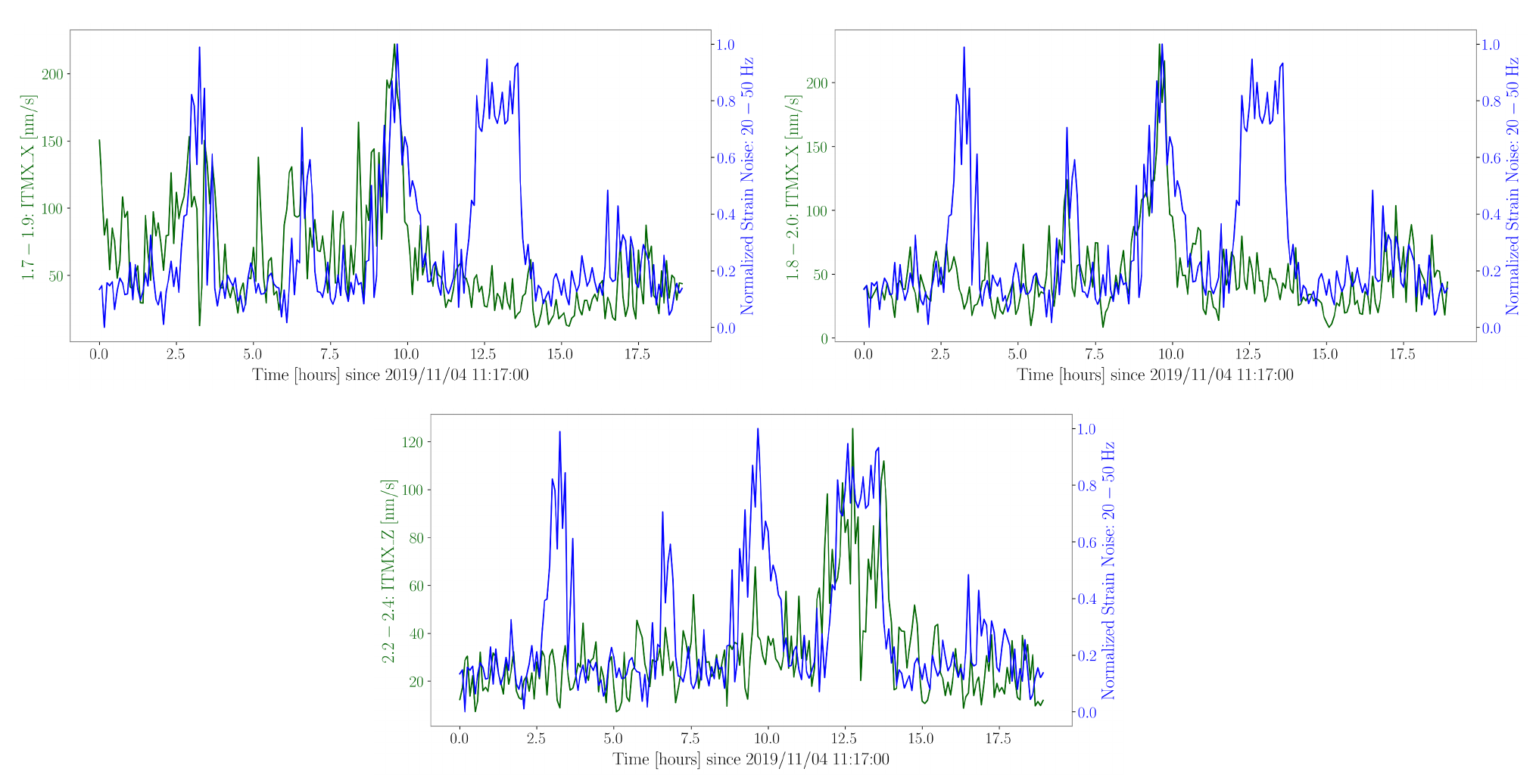}
         
    \end{subfigure}
    \caption{$1.7$--$1.9~\mathrm{Hz}$, $1.8$--$2.0~\mathrm{Hz}$, and $2.2$--$2.4~\mathrm{Hz}$ seismic band passes in ITMX\_X and ITMX\_Z compared to the 20-50 Hz strain. Different frequencies correspond to the different bursts in power, suggesting multiple scattering surfaces.}
    \label{fig:191104train}
\end{figure}

The most common frequency ranges that correlate with increases in detector noise are $0.6$--$0.8~\mathrm{Hz}$, $1.7$--$1.9~\mathrm{Hz}$, $1.8$--$2.0~\mathrm{Hz}$, and $2.3$--$2.5~\mathrm{Hz}$. Frequently seen in the spectrograms of these trains is different narrow seismic frequencies correlating with the various bursts of increased power. This can be seen in figure \ref{fig:191104train}, and suggests that there may be multiple scattering surfaces. More examples similar to figure \ref{fig:191104train} can be found in the appendix.

\newpage
\subsection{O3 results from correlation study}
Table \ref{tab:O3_trains_rate_motion} shows the results for Spearman correlation analysis between ground motion and rate of transients in the strain channel for trains in O3. The ground motion at each location and along each axis, is divided into several different frequency bands. In this analysis, we find the fast scatter rate correlates best with corner station in $1.8$--$2.0~\mathrm{Hz}$. Remarkably, as can be observed from this table, the corner station has the highest correlation with the transient rate in all the bands above $1.5$ Hz. Its also worth stressing that the corner station couples only slightly better than other locations. However, the ground motion frequency bands excited by the trains are very similar at all locations and so even small differences in coupling correlations can be meaningful. 
\begin{table}[h]
\centering
\setlength{\tabcolsep}{6pt}
    
\begin{tabular}{|c|c|c|c|c|c|c|c|c|c|} 
\hline
\multicolumn{1}{c}{Frequency band} & \multicolumn{1}{c}{IY{\_}X} & \multicolumn{1}{c}{IY{\_}Y} & \multicolumn{1}{c}{IY{\_}Z} & \multicolumn{1}{c}{EX{\_}X} & \multicolumn{1}{c}{EX{\_}Y} & \multicolumn{1}{c}{EX{\_}Z} & \multicolumn{1}{c}{EY{\_}X} &\multicolumn{1}{c}{EY{\_}Y} &\multicolumn{1}{c}{EY{\_}Z}\\ 
 \hline
0.3 - 0.6 &0.16 & \textbf{0.18}& 0.12& 0.16& 0.17&	0.12& 0.15& \textbf{0.18}&	0.11 \\
\hline
0.6 - 0.9 & 0.36	&0.42 &	0.18&	\textbf{0.49}&	0.47&	0.22&	0.36&	0.28&	0.15\\
\hline
0.9 - 1.2 & 0.28 & 0.39 & 0.31 & 0.39 & 0.37 & \textbf{0.41} & 0.18 & 0.12  &  0.27\\
\hline
1.2 - 1.5 & 0.46 & 0.53	& 0.44	&0.51	&0.51	&\textbf{0.56}	&0.26	&0.18	&0.30\\
\hline
1.5 - 1.8 & \textbf{0.66}	&0.64	&0.60	&\textbf{0.66}	&0.65	&\textbf{0.66} &	0.45	&0.37	&0.48 \\
\hline
1.8 - 2.1 & \textcolor{magenta}{\textbf{0.69}} &  \textcolor{magenta}{\textbf{0.69}} & 0.64	& 0.66	& 0.63	& 0.65	& 0.54	& 0.56	& 0.56\\
\hline
2.1 - 2.4 & 0.65 & \textbf{0.66} &0.59	&0.63	&0.60	&0.58	&0.53	&0.53	&0.52\\
\hline
2.4 - 2.7 & 0.65 & \textbf{0.67}	& 0.61 & 0.62 &	0.61 &	0.58	&0.54	&0.58	&0.51\\
\hline
2.7 - 3.0 & 0.60 & \textbf{0.61} &	\textbf{0.61}&	\textbf{0.61}&	0.59&	0.56&	0.49&	0.52&	0.47\\
\hline
3.0 - 3.3 & 0.61	& \textbf{0.62} & 	0.61& 	0.61&	0.58	&0.57	&0.54	&0.55	&0.52\\
\hline
3.3 - 3.6 & 0.59 &	\textbf{0.61}&	0.57&	0.60&	0.57&	0.56&	0.49&	0.51&	0.47\\
\hline
3.6 - 3.9 & 0.60	&\textbf{0.61}	&0.45	&0.58	&0.57	&0.56	&0.46	&0.51	&0.47\\
\hline
3.9 - 4.2 & 0.56	&\textbf{0.62}	&0.54	&0.58	&0.58	&0.56	&0.45	&0.47	&0.49\\
\hline
4.2 - 4.5 & 0.62	&\textbf{0.68}	&0.62	&0.59	&0.56	&0.52	&0.41	&0.43	&0.50\\
\hline
4.5 - 4.8 & 0.58	&\textbf{0.63}	&0.37	&0.60	&0.50	&0.51	&0.39	&0.39	&0.46\\
\hline
\end{tabular}

    \caption{Spearman correlation coefficients between band-passed ground motion data from corner, X and Y end stations seisomemeter and transient noise rate in the strain data channel during trains. The highest coefficient within each band is shown in bold. The highest correlation co-efficient across all the bands, shown in magenta, is for corner station motion in the $1.8$--$2.1~\mathrm{Hz}$  band. }
    \label{tab:O3_trains_rate_motion} 
\end{table}

\newpage
\section{Discussion}
A few conditions impact the rate of transients due to ground motion in the vicinity of the instrument. This involves the intensity of ground motion, resonant frequencies of the detector components and scattered light amplitude. As mentioned earlier, the amount of ground motion is captured by the seismometers at End X, End Y, and the corner station, and it shows a clear correlation with scattered light transients in DARM. As expected, higher ground motion leads to more noise for a given band of importance. The second key factor is the resonant vibrating motion of different detector components from which light can get scattered. Furthermore, these resonances are the reason why we have fast scatter at specific frequencies such as $2$ Hz, $4$ Hz, $3.3$ Hz. The spacing between the fast scattering arches helps us narrow down the list of possible suspects. So similar degree of ground motion in a band that contains vibration resonances will create more noise in DARM than in a band that does not. Finally, the third factor is how much light is scattered by the moving detector component, which depends on how much light is incident on it and what fraction of it is reflected towards the mirrors. This is often difficult to measure, and commissioners use different techniques, including taking photographs and videos of detector hardware to assess the amount of light on them.  Since the coupling between ground motion and strain noise depends on all these aspects, it can get very difficult to find the exact source of noise. A component with large motion amplitude may not have enough light amplitude, another component receiving sufficient light may be damped properly and thus would require too much motion to create noise above the strain sensitivity. Trains near LIGO Livingston routinely create strain noise, mostly in 10-200 Hz frequency band. The ground motion amplitude due to these trains is highest at End Y but a priori it cannot be assumed to be location of noise coupling, as per the above discussion. In this paper, we answer two main questions with regards to noise due to trains. What ground motion frequency band correlates the most with strain noise? Which location (out of End Y, End X and Corner) has highest ground motion coupling with strain noise? To answer these questions we use two methods, the Lasso analysis and the Spearman correlation analysis. Using both these approaches, we find a high degree of correlation between ground motion in $1.8$--$2.2~\mathrm{Hz}$ at the Corner station and strain noise due to trains. Identification of frequency bands that correlate well with strain noise is an important step as it allows commissioners to narrow down the list of potential suspects with resonances within the band. These resonances may then be damped leading to reduction in the noise. We suspect the motion of ARM Cavity Baffle in the corner station could be responsible for increased noise in $h(t)$ during trains. 

\newpage
\section{Appendix}
Here we provide more examples of seismic band pass plots of various trains.

\begin{figure}[!tph]
\captionsetup[subfigure]{font=scriptsize,labelfont=scriptsize}
   \centering
    \begin{subfigure}[b]{1.0\textwidth}
        \centering
         \includegraphics[width= \textwidth]{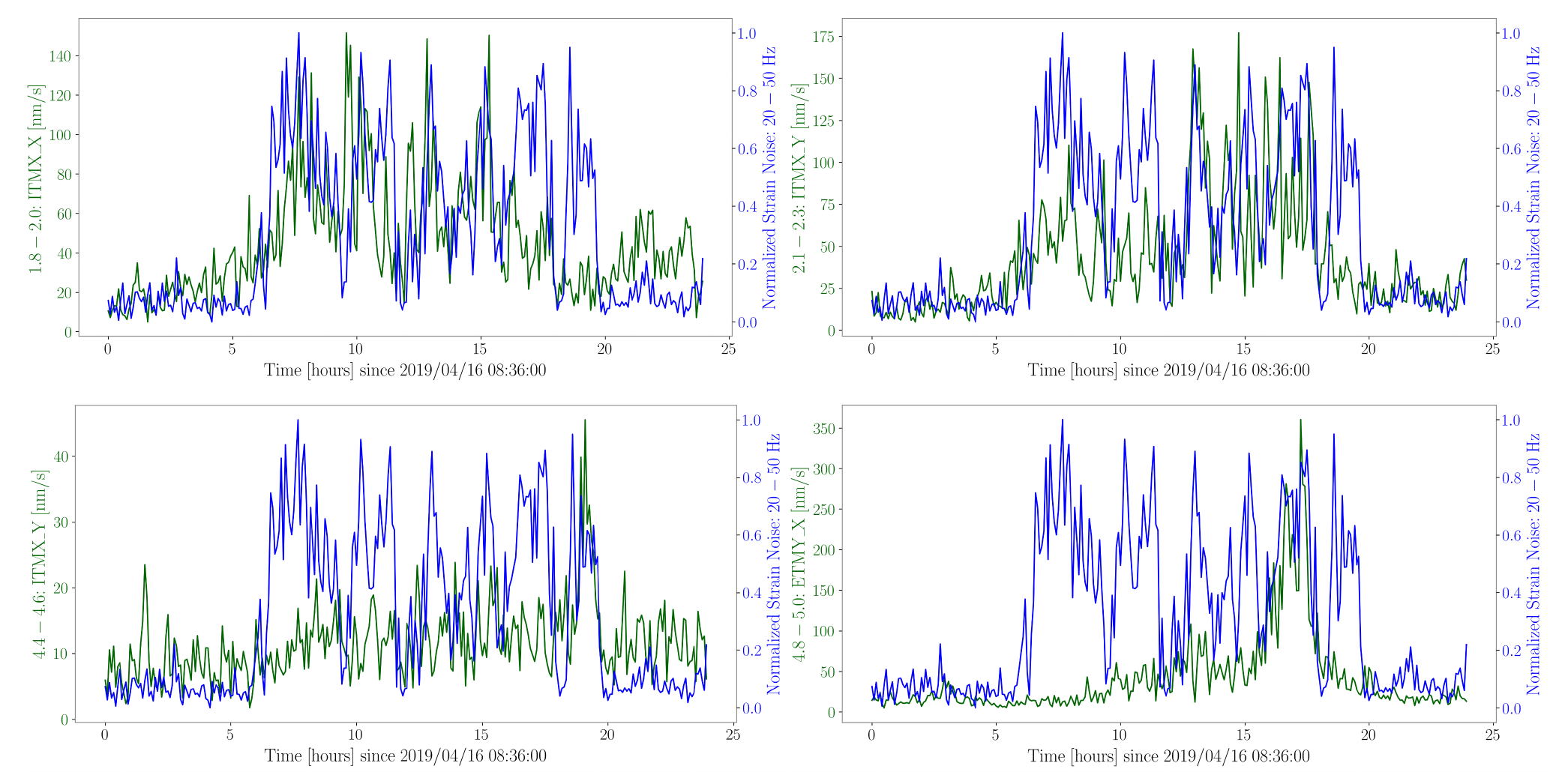}
         \label{fig:190416train}
         \caption{Train $2019$--$04$--$16$ at 08:36:00.}
    \end{subfigure}
\end{figure}

\begin{figure}[!tph]
\captionsetup[subfigure]{font=scriptsize,labelfont=scriptsize}
   \centering
    \begin{subfigure}[b]{1.0\textwidth}
        \centering
         \includegraphics[width= \textwidth]{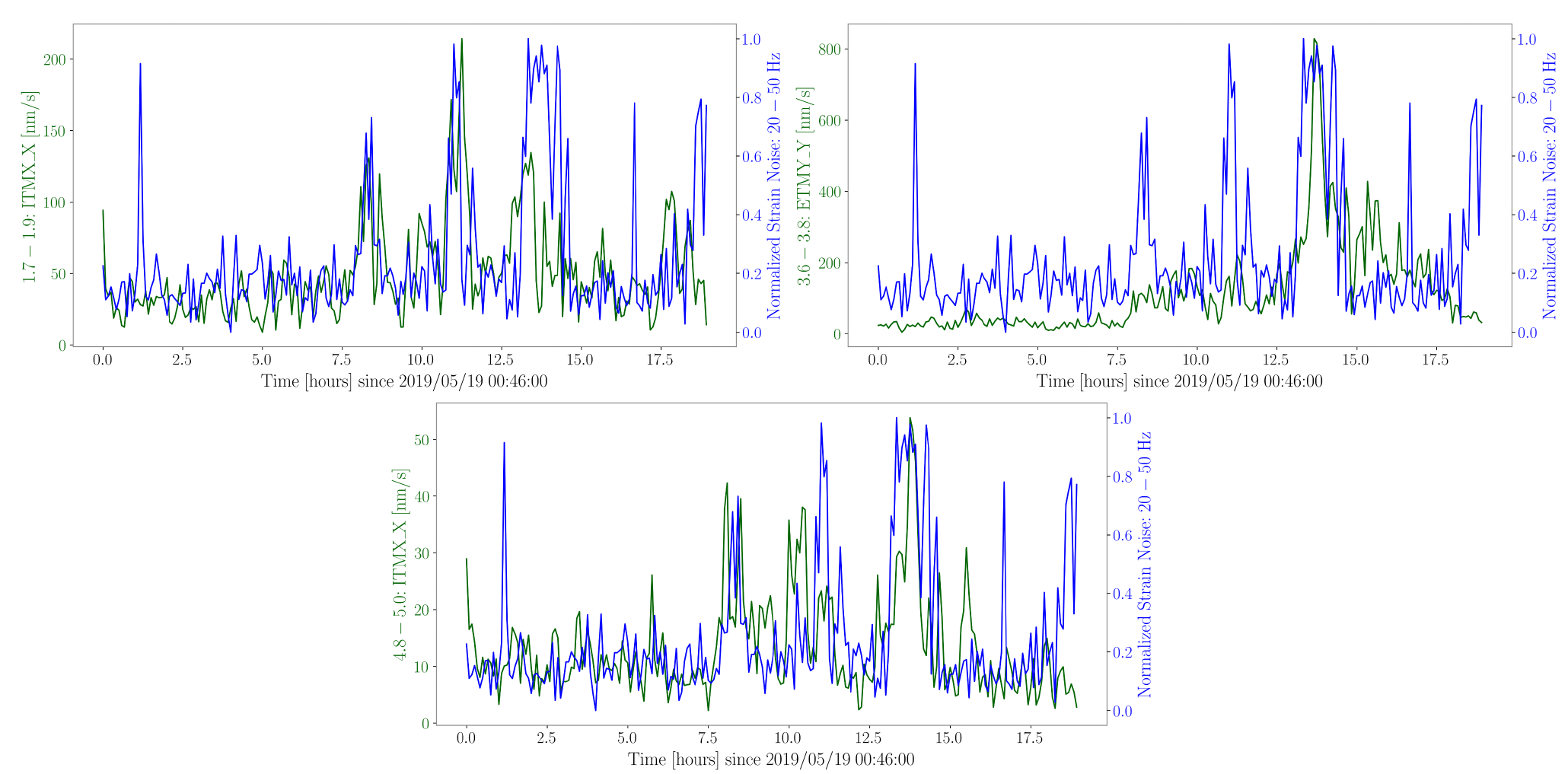}
         \label{fig:190519train}
         \caption{Train $2019$--$05$--$19$ at 00:46:00.}
    \end{subfigure}
\end{figure}

\begin{figure}[!tph]
\captionsetup[subfigure]{font=scriptsize,labelfont=scriptsize}
   \centering
    \begin{subfigure}[b]{1.0\textwidth}
        \centering
         \includegraphics[width= \textwidth]{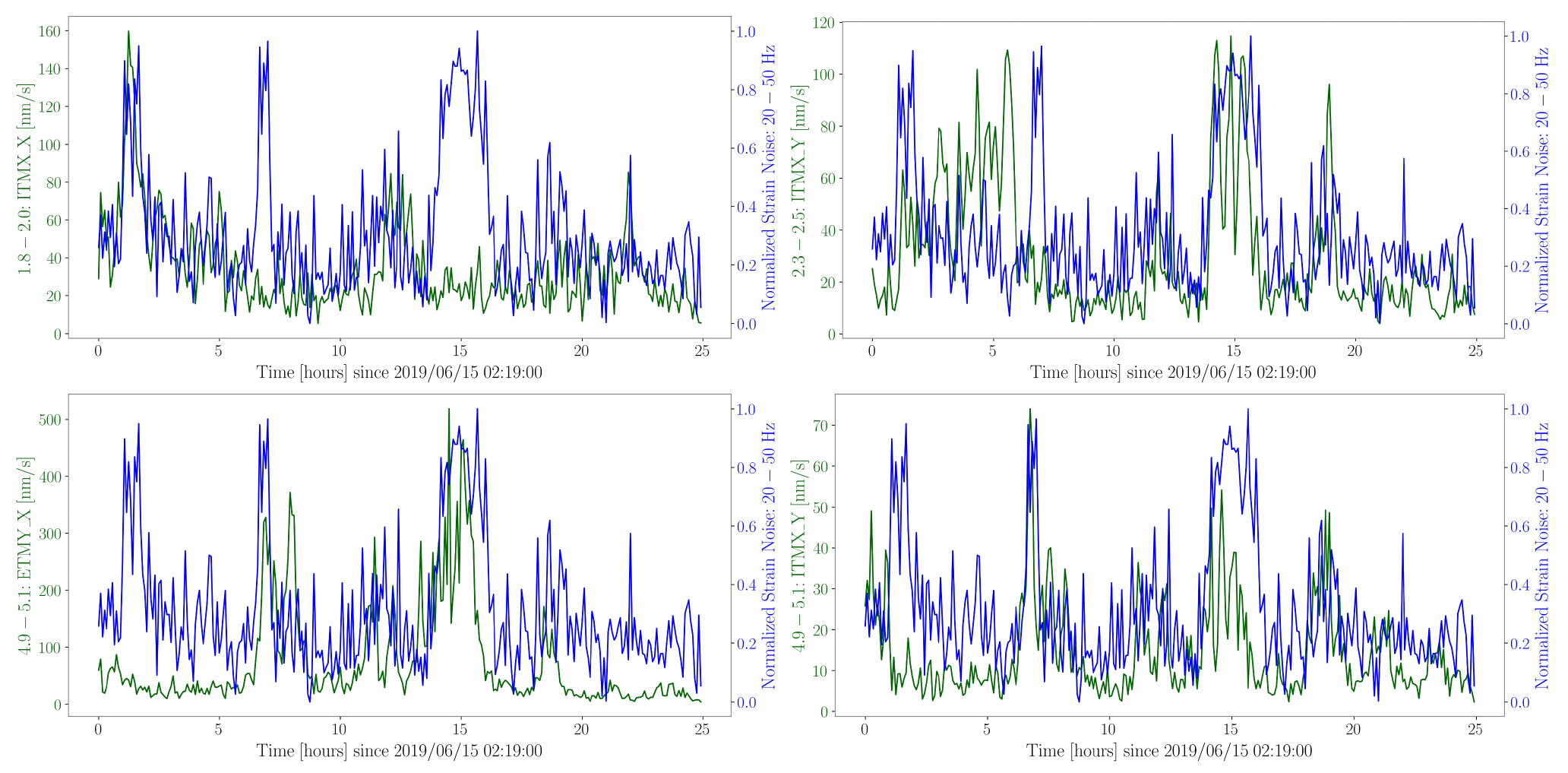}
         \label{fig:190615-2utc-train}
         \caption{Train $2019$--$06$--$15$ at 02:19:00.}
    \end{subfigure}
\end{figure}

\begin{figure}[!tph]
\captionsetup[subfigure]{font=scriptsize,labelfont=scriptsize}
   \centering
    \begin{subfigure}[b]{1.0\textwidth}
        \centering
         \includegraphics[width= \textwidth]{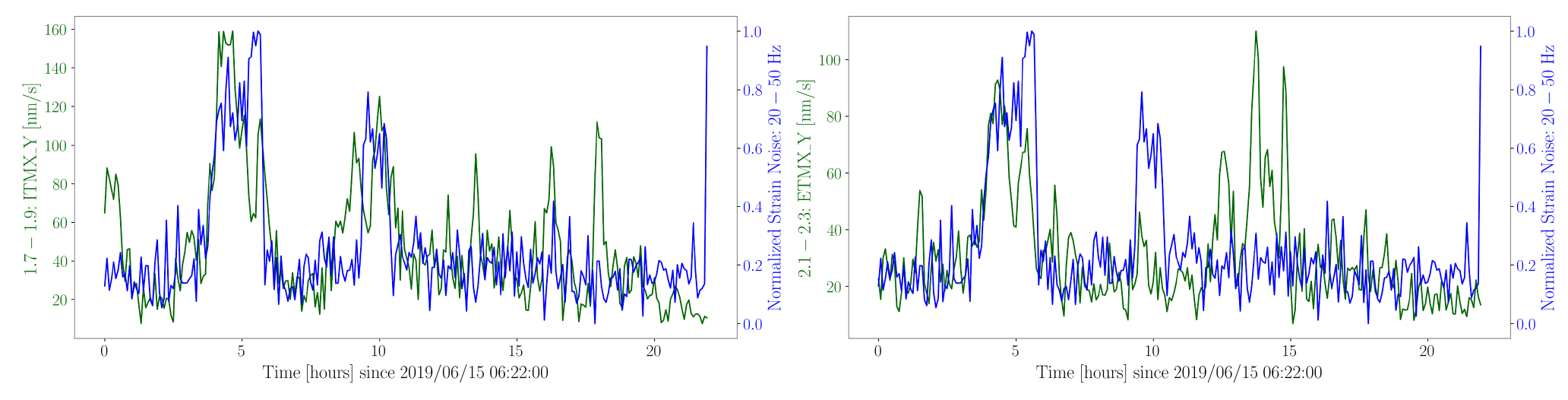}
         \label{fig:190615-6utc-train}
         \caption{Train $2019$--$06$--$15$ at 06:22:00.}
    \end{subfigure}
\end{figure}

\newpage
\bibliographystyle{iopart-num}
\bibliography{trains.bib}

\end{document}